\documentclass[journal]{IEEEtran}
\usepackage[dvipdfmx]{graphicx,color}

\usepackage[OT1]{fontenc} 
\usepackage[numbers,sort&compress]{natbib}
\usepackage[cmex10]{amsmath}
\usepackage{amssymb}
\usepackage{bm}
\usepackage[dvipdfmx]{graphicx}
\usepackage{color}
\usepackage{subfigure}
\usepackage{mathtools}
\usepackage{multirow}
\usepackage{algorithm,algorithmic}
\usepackage{listings}
\usepackage{comment}
\usepackage{pmat}
\usepackage{tikz}
\usepackage{cases}
\usepackage{hyperref}
\usepackage{breakurl}

\def\I{\mathbf{I}}
\def\E{\mathbf{E}}

\def\Y{\mathbf{Y}}

\def\Yh{\hat{\mathbf{Y}}}

\def\H{\mathbf{H}}

\def\S{\mathbf{S}}

\def\V{\mathbf{V}}

\def\X{\mathbf{X}}
\def\Xh{\hat{\mathbf{X}}}

\def\G{\mathbf{G}}
\newcommand{\argmin}{\mathop{\rm arg~min}\limits}

\makeatletter
\def\ps@IEEEtitlepagestyle{%
  \def\@oddfoot{\mycopyrightnotice}%
  \def\@oddhead{\hbox{}\@IEEEheaderstyle\leftmark\hfil\thepage}\relax
  \def\@evenhead{\@IEEEheaderstyle\thepage\hfil\leftmark\hbox{}}\relax
  \def\@evenfoot{}%
}
\def\mycopyrightnotice{%
  \begin{minipage}{\textwidth}
  \centering \scriptsize
\textcopyright 2023 IEEE. Personal use of this material is permitted.
  Permission from IEEE must be obtained for all other uses, in any current or future
  media, including reprinting/republishing this material for advertising or promotional
  purposes, creating new collective works, for resale or redistribution to servers or
  lists, or reuse of any copyrighted component of this work in other works.
  DOI: \href{https://doi.org/10.1109/LWC.2022.3233619}{10.1109/LWC.2022.3233619}
  \end{minipage}
}
\makeatother

\begin{document}
\title{A New Noncoherent Gaussian Signaling Scheme for Low Probability of Detection Communications}

\author{Yuma~Katsuki,~\IEEEmembership{Graduate~Student Member,~IEEE},
Giuseppe~Thadeu~Freitas~de~Abreu,~\IEEEmembership{Senior~Member,~IEEE},
Koji~Ishibashi,~\IEEEmembership{Senior~Member,~IEEE}, and Naoki~Ishikawa,~\IEEEmembership{Senior~Member,~IEEE}.\thanks{Y.~Katsuki and N.~Ishikawa are with the Faculty of Engineering, Yokohama National University, 240-8501 Kanagawa, Japan (e-mail: ishikawa-naoki-fr@ynu.ac.jp). G.~Abreu is with the School of Computer Science and Engineering, Jacobs University Bremen, 28759 Bremen, Germany. K.~Ishibashi is with the Advanced Wireless and Communication Research Center, The University of Electro-Communications, 182-8585 Tokyo, Japana. This work was supported in part by the Japan Science and Technology Agency, Strategic International Collaborative Research Program (JST SICORP), Japan, under Grant JPMJSC20C1.}}

\markboth{Author's final version accepted for publication in IEEE Wireless Communications Letters}
{Shell \MakeLowercase{\textit{et al.}}: Bare Demo of IEEEtran.cls for Journals}
\maketitle

\begin{abstract}
We propose a novel, Gaussian signaling mechanism for low probability of detection (LPD) communication systems with either single or multiple antennas. The new scheme is designed to allow the noncoherent detection of Gaussian-distributed signals, enabling LPD communications using signals that follow the complex Gaussian distribution in the time and frequency domains. It is demonstrated via simulations that the proposed scheme achieves better performance than a comparable conventional scheme over the entire SNR region, with the advantage becoming more significant in scenarios with lower overhead. 
\end{abstract}


\IEEEpeerreviewmaketitle

\section{Introduction}
\IEEEPARstart{L}{ow} probability of detection (LPD) systems are a new type of covert wireless communication technology \cite{bash2015hiding} that is expected to play a key role in applications such as stealth IoT and military networks \cite{yan_low_2019}, which require high levels of security.
In the LPD communication systems, a legitimate transmitter attempts to communicate with a legitimate receiver without being detected by an illegitimate adversary.

Low-power zero-mean complex-valued Gaussian signals are ideal for LPD, as it has been recently shown \cite{yan_gaussian_2019} that such signals minimize the probability of detection by illegitimate adversaries.
This is unlike most current wireless systems, many of which rely on orthogonal frequency division multiplexing (OFDM) waveforms.
This is because although time-domain OFDM signals also follow Gaussian distributions, the discrete nature of the digital constellations utilized in such systems is visible in the frequency domain, a feature which can therefore be exploited to detect the presence of communications.
In contrast, in LPD systems, signals must be Gaussian in both the time and the frequency domains.

The security of LPD communication has been analyzed from an information-theoretic perspective \cite{bendary_achieving_2021, zheng2019multiantenna}, but under the assumption that perfect channel state information (CSI) is available at both transmitter and receiver \cite{bendary_achieving_2021, zheng2019multiantenna}, which in turn implies the exchange of reference signals, thus increasing the risk of detection by an adversary.
This fundamental weakness of existing LDP methods calls for the design of noncoherent LPD schemes, which on the other hand is challenging under the optimal LDP Gaussian signaling, and therefore remains an open issue hindering the practicality of the technology \cite{yan_low_2019}.

Against this background, a noncoherent detection scheme for LDP systems employing Gaussian signals is proposed in this letter, as an enabling technology for LPD communications.
The proposed scheme builds on principles of differential encoding \cite{ishikawa_differential_2018} to construct complex Gaussian reference and data signals, such that thanks to the differential structure, no periodic reference signals are required to track CSI.
The proposed scheme has advantages over the representative Gaussian signaling scheme of \cite{okamoto_chaos_2011,okamoto_performance_2016} decoded via the semi-blind detector \cite{chen_semi-blind_2010}, which can be considered the state-of-the-art in the area.
The contributions of the article can be summarized as follows.

\begin{itemize}
\item \textbf{We design a new noncoherent detection scheme for LPD communications}. The proposed scheme inserts a reference matrix at the beginning of the transmission frame and generates differentially-encoded symbols, all of which follow the ideal Gaussian distribution in the frequency domain, resulting in Gaussian-distributed time-domain signals.
Since Gaussian-distributed signals make CSI estimation difficult due to their maximally-entropic feature, the issue is circumvented by the design of a robust noncoherent detector.
\item \textbf{We demonstrate that the proposed scheme achieves better BER performance than state-of-the-art of LPD methods employing Gaussian signaling  \cite{okamoto_chaos_2011, okamoto_performance_2016}} with semi-blind detector \cite{chen_semi-blind_2010}.
In addition, our analysis indicates that the detection complexity is reduced by a linear factor while maintaining the same security level as the ideal Gaussian signaling.
\end{itemize}

\section{System Model\label{sec:sys}}
Consider a multiple-input multiple-output (MIMO) system in which a legitimate transmitter, \textit{Alice}, equipped with $M$ transmit antennas, communicates with a legitimate receiver, \textit{Bob}, equipped with $N$ receive antennas, in the presence of an illegitimate adversary, \textit{Willie}, which tries to detect Alice's communications.
The narrow-band received signals at Bob is modeled as\footnote{We remark that this system model is readily applicable to MIMO-OFDM scenarios \cite{ishikawa_differential_2018}.} 
\begin{equation}
\Y(i) = \H(i)\S(i) + \V(i) \in \mathbb{C}^{N \times T},
\label{eq:Bob:blockmodel}
\end{equation}
for $0 < i \leq W$, where $i$ is the transmission index, $W$ is the frame length, $T$ is the number of time slots, $\S(i) \in \mathbb{C}^{M \times T}$ is a space-time codeword, and the elements of $\H(i) \in \mathbb{C}^{N \times M}$ and $\V(i) \in \mathbb{C}^{N \times T}$ follow $\mathcal{CN}(0,1)$ and $\mathcal{CN}(0,\sigma_v^2)$, respectively, with the signal-to-noise ratio (SNR) is defined as $1/\sigma_v^2$. 

Let $B$ be the number of information bits conveyed by the codeword $\S(i)\in\mathcal{S}$, where $\mathcal{S}\triangleq\{\S_1, \cdots, \S_{2^B}\}$ is the codebook.
The bit error rate (BER) associated with signals as in \eqref{eq:Bob:blockmodel} depends on the so-called coding gain 
\cite{hanzo_near-capacity_2009}
\begin{align}
G\triangleq\min_{p \neq q \in \{1, \cdots, 2^B\}} \left((\S_p-\S_q)^{\mathrm{H}}(\S_p-\S_q) \right)^\frac{1}{N},
\label{eq:coding:gain}
\end{align}
which represents the minimum Euclidean distance between different space-time codewords  in the codebook.

We emphasize that the main interest of LPD communication with $M \ll N$ is in uplink scenario, since for massive MIMO systems, downlink physical layer security schemes such as the one proposed in \cite{dean2017physicallayer} can generate Gaussian signals.
As a consequence, LPD communications in downlink will be considered out of the scope of this letter.

\section{Reference State-of-the-Art Schemes}
Before we introduce our proposed method, let us briefly revise relevant state-of-the-art techniques, in particular the conventional Gaussian signaling scheme of
\cite{okamoto_chaos_2011,okamoto_performance_2016} and the conventional semi-blind detector of \cite{chen_semi-blind_2010}.
Combining both schemes is a straightforward task, and results in an exclusive Gaussian signaling scheme that can work for MIMO-OFDM scenarios.
Hence, we regard the combination as a performance baseline reference to our contribution.

\subsection{Conventional Gaussian Signaling Scheme of 
\cite{okamoto_chaos_2011,okamoto_performance_2016}}
\label{subsec:C:MIMO}

Chaos MIMO (C-MIMO) \cite{okamoto_chaos_2011} is a MIMO modulation scheme designed with basis on chaos theory\footnote{In our simulations, C-MIMO achieved the best coding gain among physical layer encryption schemes. Thus, we consider it to be a representative Gaussian signaling scheme.}, in which $B=MT$ information bits $\mathbf{b}=[b_1~b_2~\cdots~b_B]\in\mathbb{B}^{B}$ are mapped onto complex-valued Gaussian symbols, such that the associated the $M \times T$ space-time codeword is in a form \cite{okamoto_performance_2016}
\begin{equation}
\S(i)\triangleq\frac{1}{\sqrt{M}}\!\left[\begin{array}{cccc}
	s_1 & s_{M+1} & \cdots & s_{MT-M+1}\\
	s_2 & s_{M+2} & \cdots & s_{MT-M+2}\\
	\vdots & \vdots & \ddots & \vdots\\
	s_M& s_{2M} & \cdots  &s_{MT}
	\end{array}\right]\!\in \mathbb{C}^{M \times T},
	\label{eq:def:chaos}
\end{equation}
where each symbol $s_k$ is generated by the Box-Muller's transform \cite{okamoto_performance_2016}
\begin{equation}
s_k = \sqrt{-\log \left( c_k^{(x)} \right)}
\left(
    \cos \left( 2 \pi c_k^{(y)} \right) + j
    \sin \left( 2 \pi c_k^{(y)} \right)
\right)
\label{eq:box:mu}
\end{equation}
and $j$ denotes the imaginary number.

Here, $c_k^{(x)}$ and $c_k^{(y)}$ are uniform pseudo-random numbers generated by a shared key, the input bit sequence, and the Bernoulli shift map transition,
which are given by \cite{okamoto_performance_2016}
\begin{subequations}
\begin{eqnarray}
&c_k^{(x)} = \mathrm{arccos}\left(\mathrm{cos}\left(37\pi\left(\mathrm{Re}[c_k]+\mathrm{Im}[c_k]\right)\right)\right)/\pi,&\\
&c_k^{(y)} = \mathrm{arcsin}\left(\mathrm{sin}\left(43\pi\left(\mathrm{Re}[c_k]\!-\!\mathrm{Im}[c_k]\right)\right)\right)/\pi\!+\!1/2,&
\end{eqnarray}
\end{subequations}
where we have
\begin{equation}
    	c_k=\mathrm{Re}[z_{N_s+b_{(k+B/2)~\mathrm{mod}~B}}]+\mathrm{Im}[z_{N_s+b_{(k+B/2+1)~\mathrm{mod}~B}}].
\label{eq:chaosk}
\end{equation}

In \eqref{eq:chaosk}, the chaos sequences are defined by 
\cite{okamoto_performance_2016}
\begin{subequations}
\begin{eqnarray}
&\mathrm{Re}[z_{l}]=2\cdot\mathrm{Re}[z_{l-1}]~\mathrm{mod}~(1-10^{-16}),&\\
&\mathrm{Im}[z_{l}]=2\cdot\mathrm{Im}[z_{l-1}]~\mathrm{mod}~(1-10^{-16}).&
\end{eqnarray}
\end{subequations}
for $l=1,2,\cdots, N_s, N_s+1$, where the transition factor is typically set to a large number such as $N_s=100$ \cite{okamoto_performance_2016}, and both sequences are initialized by $\mathrm{Re}[z_{0}]=\Gamma(\mathrm{Re}[c_{k-1}],b_{k-1})$, $\mathrm{Im}[z_{0}]=\Gamma(\mathrm{Im}[c_{k-1}],b_{k~\mathrm{mod}~B})$, where \cite{okamoto_performance_2016}
\begin{align}
\Gamma(a,b)\triangleq\left\{
\begin{array}{ll}
a & \text{if}\;(b=0), \\
1-a & \text{if}\;(b=1~\mathrm{and}~a>1/2),\\
a+1/2 & \text{if}\;(b=1~\mathrm{and}~a\leq 1/2),
\end{array}
\right.
\end{align}
where $c_0\in\mathbb{C}$ is a pre-shared key that satisfies $0<\mathrm{Re}[c_0]<1$ and $0<\mathrm{Im}[c_0]<1$.

We remark that due to the Box-Muller transform described by \eqref{eq:box:mu}, the resultant symbol $s_k$ follows $\mathcal{CN}(0,1)$.\footnote{The open-source implementation of C-MIMO is available at \url{https://github.com/ishikawalab/wiphy/blob/master/wiphy/examples/okamoto2012chaos.py}}

\subsection{Conventional Coherent and Semi-Blind Detection of \cite{chen_semi-blind_2010}\label{subsec:conv:dec}}
\begin{algorithm}[h]
\caption{Conventional Semi-blind Detection \cite{chen_semi-blind_2010}. \label{alg:semi}}
\begin{algorithmic}[1]
    \renewcommand{\algorithmicrequire}{\textbf{Input:}}
    \renewcommand{\algorithmicensure}{\textbf{Output:}}
    \REQUIRE $\bar{\Y} =
    \begin{pmat}[{|||}]
    \Y(1) & \Y(2) & \cdots  & \Y(W)\cr
    \end{pmat}\in \mathbb{C}^{N \times WT},~ \hat{\H}^{(0)}$
    \ENSURE $\hat{\S}^{(l)}$
     \STATE {Set the iteration index $l=0$}.
    \REPEAT
    \STATE Given $\hat{\H}^{(l)}$, perform ML detection for each sub-matrix of $\bar{\Y}$ and obtain a set of estimates
       $$\hat{\S}^{(l)} =
        \begin{pmat}[{|||}]
        \hat{\S}(1) & \hat{\S}(2) & \cdots  & \hat{\S}(W)\cr
        \end{pmat}\in \mathbb{C}^{M \times WT}$$
    \vspace{-3ex}
    \STATE Update the channel matrix by $\hat{\H}^{(l+1)} = \bar{\Y} \left[\hat{\S}^{(l)}\right]^+$.
    \STATE{Set $l=l+1$}.
    \UNTIL{$l<I_{\mathrm{max}}$}.
\end{algorithmic} 
\end{algorithm}
Under the assumption that a perfect estimate of $\hat{\H}$ is available at the receiver, maximum likelihood (ML) detection can be carried out via
\begin{equation}
\hat{\S}(i)=\argmin_{\S} \|\Y(i)-\hat{\H}\S\|^2_{\mathrm{F}},
\label{eq:ML:chaos}
\end{equation}
where an optimal $\S$ is searched over the set of C-MIMO codewords.

In practice, however, frequent transmission of reference signals is required to improve the accuracy of CSI estimation by Bob, which in turn also increases the probability that transmissions are detected by Willie.
To alleviate this problem, the semi-blind detection scheme of \cite{chen_semi-blind_2010} may be exploited.
In such a scheme, a reference signal $\I_M$ is transmited at the first instance ($i.e.$, when $i=0$), which is used to obtain a rough initial channel estimate $\hat{\H}^{(0)}$.
Then, over the subsequent transmission of $W$ blocks, the received signals are concatenated and more accurate CSI is obtained iteratively, as summarized in Algorithm~\ref{alg:semi}, where we use the pseudo-inverse matrix $\mathbf{A}^{+} = \mathbf{A}^{\mathrm{H}}\left(\mathbf{A} \mathbf{A}^{\mathrm{H}}\right)^{-1}$.

\section{Proposed Noncoherent Gaussian Signaling (NGS) for LPD Communications}
\label{sec:prop:gauss}

In this section, we introduce the proposed NGS scheme satisfying the requirements of LPD communications.

\subsection{Gaussian Signal Transmission}
\label{subsec:prop:tx}

Based on a shared secret seed, Alice and Bob generate a Gaussian reference matrix $\G \in \mathbb{C}^{M \times KM}$, where $K$ is a repetition number, and each element of $\G$ follows $\mathcal{CN}(0, 1/M)$, $i.e.$, $\mathrm{E}[\|\G\|^2_{\mathrm{F}}]=KM$.
Similarly, Alice and Bob generate a Gaussian projection matrix $\E(i) \in \mathbb{C}^{M \times T}$, where each element of $\E(i)$ follows $\mathcal{CN}(0, 1/M)$, $i.e.$, $\mathrm{E}[\|\E(i)\|^2_{\mathrm{F}}]=T$, where the index $i$ indicates that the projection matrix varies over time.

After the above Gaussian matrices are prepared, at the index $i=0$, the Gaussian reference matrix is transmitted.
Then, for $0< i \leq W$, $B$ bits of information are mapped onto a unitary matrix $\X(i) \in \mathbb{C}^{M \times M}$, which is selected from a codebook of $2^B$ matrices $\{\X_1, \cdots, \X_{2^B}\}$.
Here, we use the classic differential encoding in the form
\begin{equation}
\tilde{\S}(i) = \left\{ \begin{array}{ll}
\I_M & \text{if } i=0, \text{or}\\
\tilde{\S}(i-1)\X(i) & \text{if } i > 0.
\end{array} \right.
\label{eq:summ:differencial}
\end{equation}

Finally, the space-time codeword at $i$ is generated by
\begin{equation}
\S(i) = \left\{ \begin{array}{ll}
\G & \text{if } i=0, \text{or}\\
\tilde{\S}(i) \E(i) & \text{if } i > 0.
\end{array} \right. \label{eq:propS}
\end{equation}

The important question here is what type of unitary matrix $\X(i)$ would lead each element of \eqref{eq:propS} to follow an independent complex Gaussian distribution.
Since any point on the unit circle does not change the statistical property, one can expect that any unitary matrix that has one unit-norm nonzero element in each column results in the Gaussian distribution.
Differential schemes that satisfy such property include the diagonal unitary coding (DUC) \cite{hochwald_differential_2000} and  differential spatial modulation \cite{bian2015differential} methods, but since the DUC approach maximizes the coding gain due to its optimized structure, it will be adopted here in our NGS scheme.

In the DUC scheme \cite{hochwald_differential_2000}, the $B$ input bits are mapped onto the integers $b=0,1, \cdots, 2^B-1$, and the corresponding unitary matrix is generated as
\begin{equation}
\X_b = \mathrm{diag}\left[\mathrm{exp}\left(j\frac{2\pi b}{2^B}u_1\right), \cdots, \mathrm{exp}\left(j\frac{2\pi b}{2^B}u_M\right)\right],
\label{eq:DUC:datamatrix}
\end{equation}
where the factors $0 < u_1 \leq \cdots u_M \leq2^B/2 \in \mathbb{Z}$ are designed so as to maximize the diversity product given by\footnote{The optimized factors are available online at \url{https://github.com/ishikawalab/wiphy/blob/master/wiphy/code/duc.py}.}
\begin{equation}
\min_{b\in\{1, \cdots, 2^B-1\}}\left|\prod_{m=1}^M \mathrm{sin}\left(\frac{\pi b u_m}{2^B}\right)\right|^\frac{1}{M}\!\!\!\!\!.
\label{eq:DUC:fanc}
\end{equation}

Notice that the encoding scheme described by \eqref{eq:DUC:datamatrix} becomes identical to the classic differential phase-shift keying modulation if $M=1$ and $u_1=1$.
In addition, although the ideal transmission rate is $R=B / T$,
since the Gaussian reference matrix $\mathbf{G}^{M \times KM}$ is inserted, the effective transmission rate is

\begin{equation}
R_{\mathrm{eff}} = \frac{BW}{MK+WT} = \eta R,
\end{equation}
where we have implicitly defined the transmission efficiency $\eta \triangleq (1 + MK / (WT))^{-1}.$
%

As given, the efficiency decreases as $K$ increases.
The reference insertion ratio is calculated as $1 - \eta$.
In the performance comparisons, we will set the ratio to 5\%, but we remark that the performance of the proposed scheme remains constant in high-speed mobile scenarios even with lower reference insertion ratios such as $1\%$ and $0.1\%$.

\subsection{Noncoherent Detection of Gaussian Signaling}
\label{subsec:prop:rx}

The noncoherent detection of the Gaussian signaling scheme proposed in Subsection~\ref{subsec:prop:tx} is not a straightforward task.
We extend the noncoherent detection scheme proposed in \cite{ishikawa_differential_2018} to support time-varying Gaussian reference and projection matrices.
Specifically, for the data blocks $0 < i \leq W$, the proposed noncoherent ML detector is described by
\begin{equation}
\Xh(i)=\argmin_{\X} \|\Y(i)-\Yh(i-1)\X\E(i)\|^2_{\mathrm{F}},
\label{eq:summ:nonsq:detection}
\end{equation}
where the matrices $\Yh(i)$ are given by
\begin{equation}
\Yh(0) \triangleq \Y(0) \G^+ = \H(0) \tilde{\S}(0) + \V(0) \G^+,
\label{eq:y:hat:0}
\end{equation}
at $i=0$, where each element of $\V(0) \in \mathbb{C}^{N \times KM}$ follows $\mathcal{CN}(0,\sigma_v^2)$, and
\begin{equation}
\Yh(i) \triangleq
\beta \Y(i) \E^{\mathrm{H}}(i)\! +\!
\Yh(i-1) \Xh(i) (\I_M\! -\! \beta \E(i) \E^{\mathrm{H}}(i)),
\end{equation}
for $i > 0$, respectively, with the parameter $\beta \triangleq 1 - \alpha$ containing a forgetting factor $0 < \alpha < 1$ that determines the inter-dependence between $\Y(i)$ and $\Yh(i-1)$. 

Notice that due to \eqref{eq:summ:nonsq:detection}, even if Willie successfully detects the LPD communications, he cannot decrypt data symbols because the projection of $\E(i)$ induces phase ambiguity, which implies that the proposed NGS also works as a physical layer encryption scheme.

\section{Theoretical Analysis}

In this section, we compare the proposed NGS LPD communication scheme against the conventional C-MIMO in terms of security and complexity. 
System configurations of the conventional C-MIMO and the proposed NGS schemes are summarized in Table \ref{tab:designs}.

\begin{table*}[t]
\centering

\caption{System configurations of the conventional C-MIMO and the proposed NGS schemes.}
\vspace{-1ex}
\begin{tabular}{|c||c|c|c|}\hline
  & Reference Signal ($i=0$)& Data Symbols ($0<i\leq W$)& ML Detector\\\hline
Conventional C-MIMO& Gaussian reference matrix $\G \in \mathbb{C}^{M \times KM}$& Okamoto's C-MIMO encoding method 
\cite{okamoto_performance_2016}& Semi-blind detector
\cite{chen_semi-blind_2010}\\
 & (see Section~\ref{subsec:prop:tx})& (see Section~\ref{subsec:C:MIMO})& (see Algorithm \ref{alg:semi})\\\hline
Proposed NGS & Gaussian reference matrix $\G \in \mathbb{C}^{M \times KM}$ & DUC \& Gaussian projection matrix $\E(i)$ & Noncoherent detector \\
 & (see~Section \ref{subsec:prop:tx}) & (see~Section \ref{subsec:prop:tx}) & (see~Section \ref{subsec:prop:rx})\\\hline
\end{tabular}
\label{tab:designs}
\vspace{-2ex}
\end{table*}

\subsection{Security Analysis}

In LPD communications, the achievable security level has been evaluated by the Willie's minimum detection error probability 
\cite{bendary_achieving_2021, zheng2019multiantenna, yan_gaussian_2019}.
Let the case when Alice does not transmit symbols to Bob be referred to as the null-hypothesis $\mathcal{H}_0$, with likelihood function $p_0(y)$, and the alternative case when Alice does transmit to Bob be referred to as the alternative hypothesis $\mathcal{H}_1$, with likelihood function $p_1(y)$.
The received signal at Willie under both hypotheses can be expressed as \cite{yan_gaussian_2019}
\begin{equation}
\left\{ \begin{array}{ll}
\mathcal{H}_0:y=v,\;\text{or}\\
\mathcal{H}_1:y=s+v,
\end{array} \right.
\label{eq:hypo}
\end{equation}

\begin{figure}[tb]
\centering
\includegraphics[width=\columnwidth]{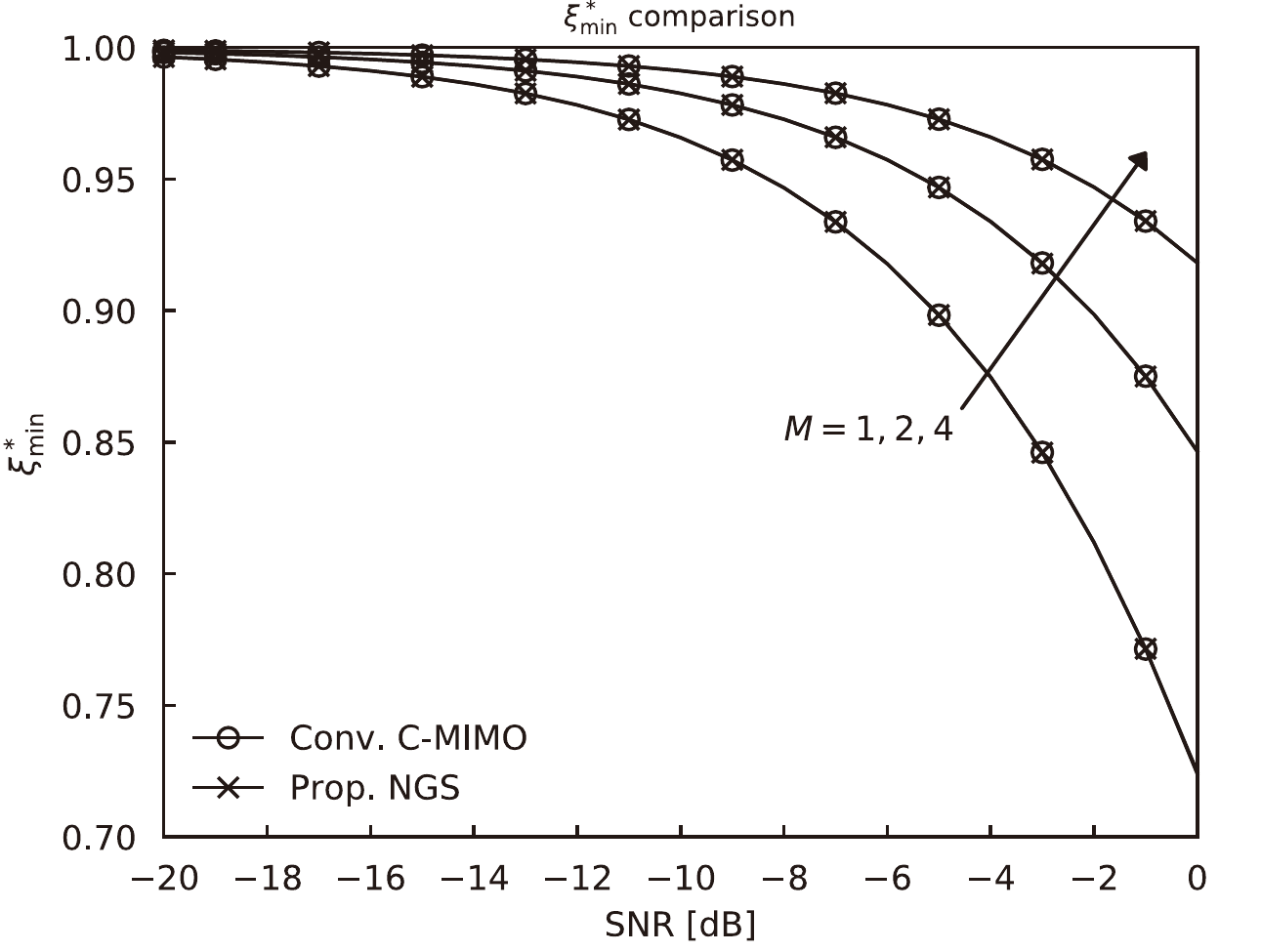}
\vspace{-4ex}
\caption{Lower bounds of Willie's minimum detection error probability $\xi^*$.}
\label{fig:KL}
\vspace{-3ex}
\end{figure}

\noindent where $s$ and $v$ denote the transmit signal and additive white Gaussian noise, respectively.

Assuming that all transmit signals are Gaussian, Willie's detection error probability can be expressed as \cite{yan_gaussian_2019}
\begin{equation}
\xi^* \geq \xi^*_{\mathrm{min}}\triangleq 1-\sqrt{\mathcal{D}(p_1\|p_0)/2},
\end{equation}
where the lower-bounding quantity $\xi^*_{\mathrm{min}}$ was derived in \cite{bendary_achieving_2021, yan_gaussian_2019}  and $\mathcal{D}(p_1\|p_0)$ denotes the Kullback-Leiber (KL) divergence between the likelihoods of observing $y$ under the null and the alternative hypotheses, respectively. 

In Fig.~\ref{fig:KL}, we evaluate the lower bounds of Willie's minimum detection error probability $\xi^*_{\mathrm{min}}$, where the conventional C-MIMO and the proposed NGS were considered.
As expected, both schemes exhibit the same lower bound since the resulting constellation in both successfully follow the ideal Gaussian distribution.
Additionally, it is observed that as $M$ increases, $\xi^*_{\mathrm{min}}$ becomes larger because less power is allocated to each antenna, which is preferable in LPD communications.
In short, it can be said that the proposed scheme has the same security level of
conventional C-MIMO.

We remark that the repetition number $K$ has no effect on $\xi^*_{\mathrm{min}}$ since it does not influence the statistical property of transmit signals, which can be inferred from \cite{bendary_achieving_2021, zheng2019multiantenna, yan_gaussian_2019}.

\subsection{Complexity Analysis}
Next, the conventional C-MIMO and the proposed NGS are evaluated in terms of encoding and decoding complexities.

\subsubsection*{Encoding complexity}
In both the C-MIMO and NGS schemes a total of $BW$ [bits] are transmitted within the frame length $W$.
However, in order to generate a codebook, C-MIMO requires $2^{B} W$ complex-valued random variables, since C-MIMO directly maps information bits onto complex-valued Gaussian symbols.

In contrast, NGS requires only $BW$ complex-valued random variables since the codebook is constructed as in \eqref{eq:DUC:datamatrix}, before being transformed by the Gaussian projection matrix.
Thus, the number of Gaussian variables to be generated can be significantly reduced in the proposed NGS LPD scheme as the number of bits $B$ increases.

\subsubsection*{Decoding complexity}

The complexity orders of the conventional semi-blind detector $C_{\mathrm{c}}$
(see Algorithm \ref{alg:semi})
and of the proposed detector $C_{\mathrm{p}}$ (see Subsection \ref{subsec:prop:rx}) are, respectively
\begin{align}
C_{\mathrm{c}}=&2^BWI_{\mathrm{max}}(\underbrace{4MNT}_{\hat{\H}\S}+\underbrace{4NT}_{\|\cdot\|^2_{\mathrm{F}}})\nonumber\\
&+\underbrace{MW(I_\mathrm{max}-1) \left(8MT+\frac{4N}{W}+\frac{M^2}{W} \right)}_{\bar{\Y} \left[\hat{\S}^{(l)}\right]^+}\nonumber\\
=&\mathcal{O}\big(2^BWI_{\mathrm{max}}(4MNT+4NT)\big),\\
C_{\mathrm{p}}=&2^BW(\underbrace{4MNT}_{\hat{\Y}(i-1)\X}+\underbrace{4NT}_{\|\cdot\|^2_{\mathrm{F}}}+\underbrace{4MT} _{\X\E(i)})\nonumber\\
&+\underbrace{MW(8MN+4NT+4MT)}_{\beta \Y(i) \E^{\mathrm{H}}(i)\! +\!
\Yh(i-1) \Xh(i) (\I_M\! -\! \beta \E(i) \E^{\mathrm{H}}(i))}\nonumber\\
=&\mathcal{O}\big(2^BW(4MNT+4NT+4MT)\big),
\end{align}
where only the significant numbers of real-valued floating-point multiplication operations are counted.

It follows that the ratio between both is given by
\begin{align}
\frac{C_{\mathrm{p}}}{C_{\mathrm{c}}}&= \mathcal{O}\left(\frac{1}{I_{\mathrm{max}}}
\left(
\frac{1 + \frac{1}{M} + \frac{1}{N}}{1 + \frac{1}{M}}
\right)\right)
\approx \mathcal{O}\left(\frac{1}{I_{\mathrm{max}}}\right),
\end{align}
which means that the complexity order of the conventional semi-blind detector is approximately $I_{\mathrm{max}}$ higher than that of the proposed detector.

\section{Numerical Performance Comparisons}
\label{sec:comp}

In this section, the conventional C-MIMO decoded via the semi-blind detector and the proposed NGS LPD scheme are empirically evaluated in terms of their bit error rates (BERs) and coding gains.
Again, the system configurations of both schemes are summarized in Table \ref{tab:designs}.
In our simulations, we set the minimum SNR to $-20$ dB since transmission power in LPD communications systems is typically small, and adopt the forgetting factor $\alpha=0.8$.
In addition, although the proposed NGS supports time-varying channels, we limit the comparisons to quasi-static Rayleigh fading channels because the conventional C-MIMO with the semi-blind detector only supports such channel condition.

Note that also in the conventional C-MIMO scheme, a Gaussian reference matrix $\G \in \mathbb{C}^{M \times KM}$ is transmitted first for the purpose of channel estimation, since all signals in LPD systems must be complex Gaussian. The semi-blind detection algorithm is subsequently performed using the rough initial channel estimate $\hat{\H}^{(0)}=\hat{\Y}(0)$ given by \eqref{eq:y:hat:0}.

\begin{figure}[tb]
\centering
\includegraphics*[width=\columnwidth]{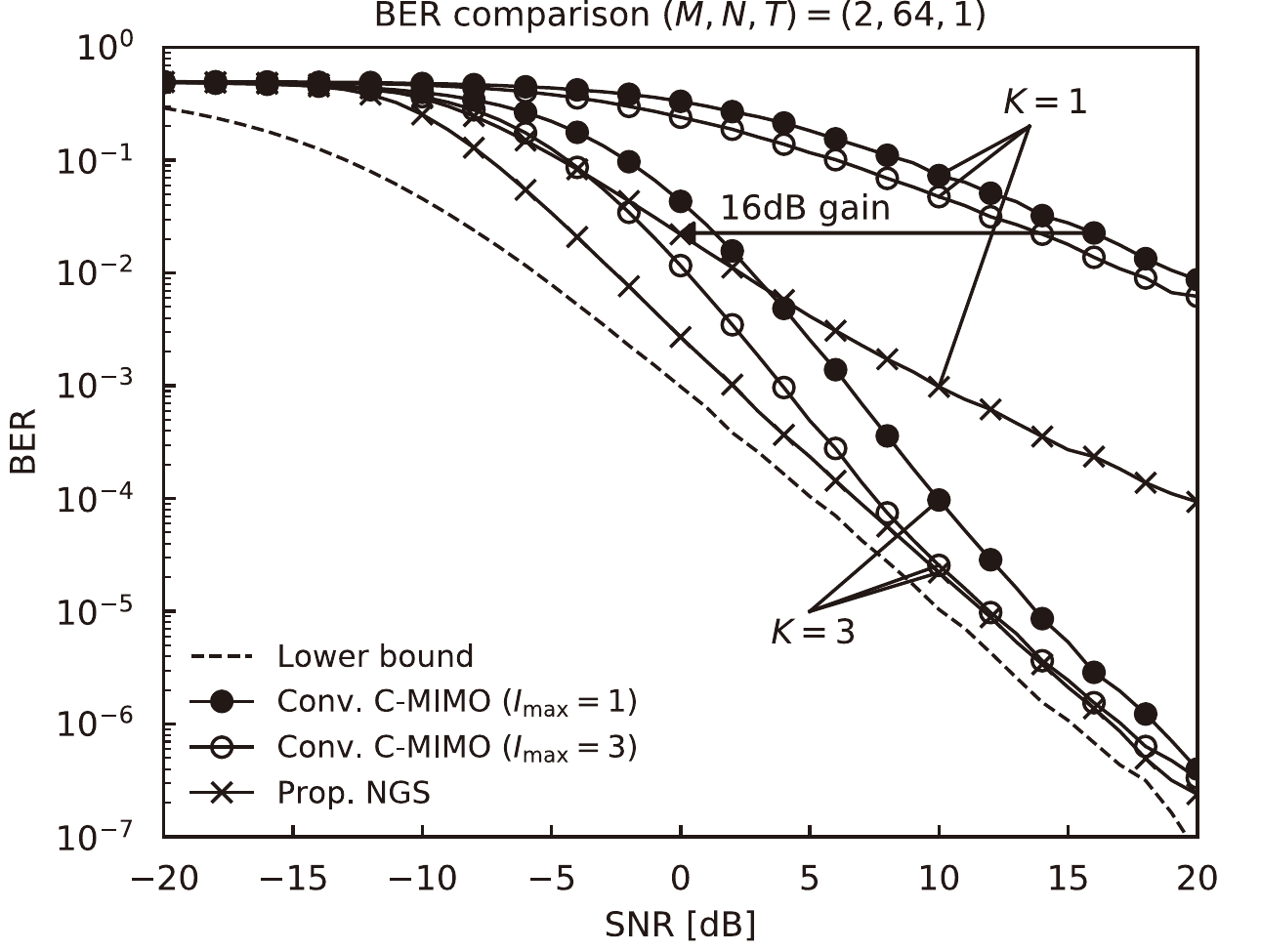}
\vspace{-4ex}
\caption{Comparison of BERs achieved with Gaussian reference signals and Gaussian projection matrix, with $M=2$, $N=64$, $T=1$ and $B=2$.}
\label{fig:comp:semi:BER:M:2}
\vspace{2ex}
\includegraphics*[width=\columnwidth]{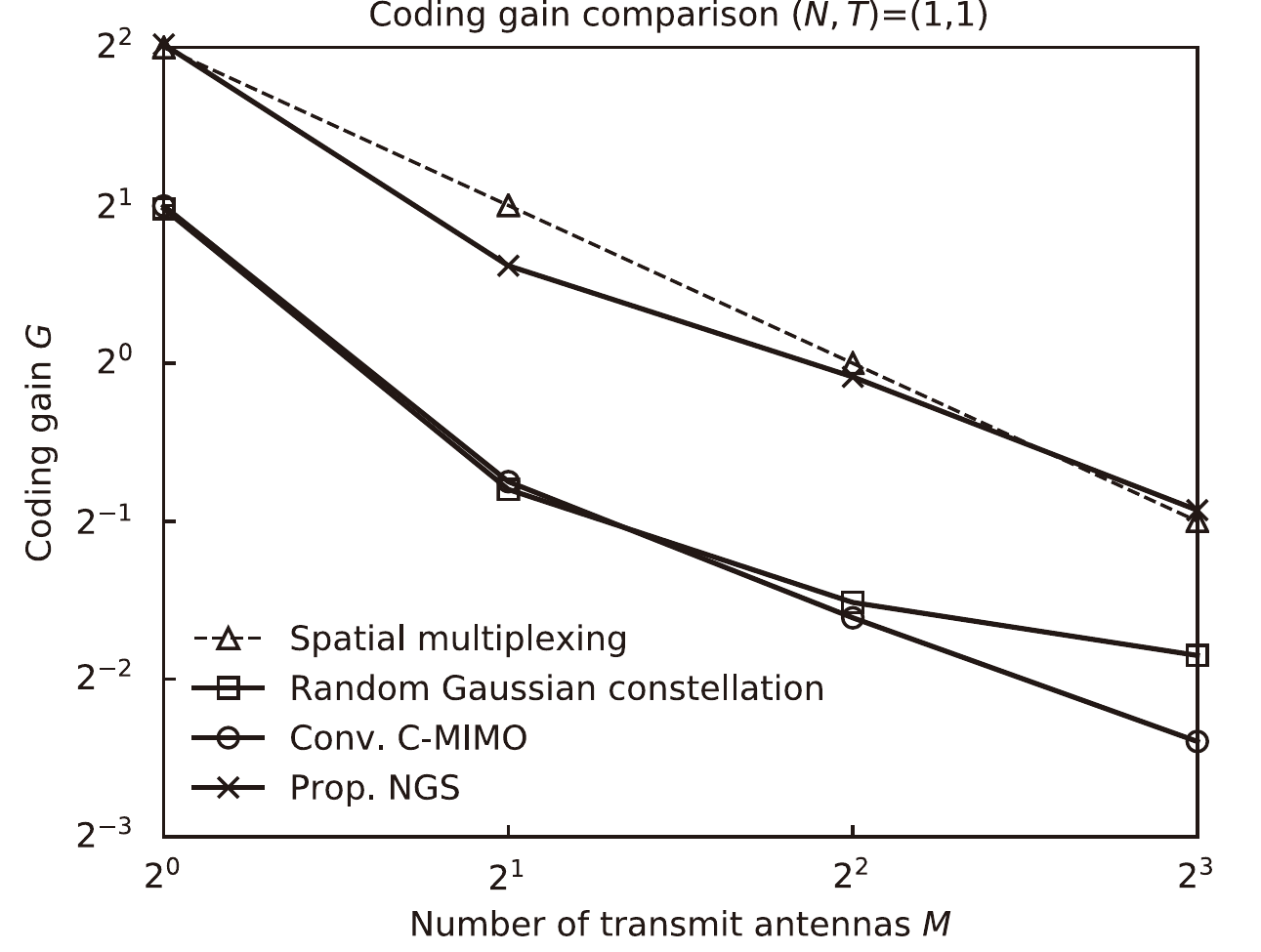}
\vspace{-4ex}
\caption{Comparison of coding gains achieved with Gaussian projection matrix, with $B=M$, $N=1$ and $T=1$.}
\label{fig:comp:coding:gain}
\vspace{-2ex}
\end{figure}

First, a comparison between the BERs of the conventional C-MIMO \cite{okamoto_chaos_2011} scheme with semi-blind detection \cite{chen_semi-blind_2010} and the proposed NGS LPD method is offered in Fig.~\ref{fig:comp:semi:BER:M:2}, for the case where $M=2$, $N=64$, $T=1$, $B=2$, and overhead factor $K=1$ or $K=3$.
As a lower bound, the BER of the NGS scheme with perfect CSI is also included.
The results show that the proposed NGS LPD system achieves the best BER in all cases, with a gain of 16 dB over conventional C-MIMO observed for $K=1$, and a performance rapidly approaching the lower bound for larger $K$.

In order to better investigate the reason behind the large BER advantage of NGS LPD seen in Fig.~\ref{fig:comp:semi:BER:M:2}, we compare in Fig.~\ref{fig:comp:coding:gain} the coding gains of NGS LPD and alternative schemes\footnotemark, with $N=1$ and $T=1$, and $M$ varying from $1$ to $8$.

\footnotetext {Note that although spatial multiplexing does not generate Gaussian signals, it was added to the comparison only as a reference to coding gain.}

The results of Fig.~\ref{fig:comp:coding:gain} show that thanks to the optimized DUC constellation, the proposed NGS LPD scheme achieves coding gains equal or close to those of conventional spatial multiplexing, while generating Gaussian signals.
In comparison, the coding gains of conventional C-MIMO was found to be close or equal to these of random Gaussian constellations.

\section{Conclusion}
\label{sec:conc}

We proposed a solution to an open issue regarding the noncoherent detection of Gaussian-distributed signals, which is especially important for LPD communications.
To solve the issue, we relied on newly proposed Gaussian projection scheme, applied over
optimized DUC constellations.
Reference signals were also designed so as to follow the Gaussian distribution.
Thus, the proposed scheme communicates only using Gaussian signals, which satisfies the common requirement of LPD communication.
With the use of OFDM, the Gaussian signals in the frequency domain result in Gaussian-distributed time-domain signals.
An analysis was provided, which clarified that the proposed NGS generates perfectly Gaussian symbols at a fraction of the complexity of C-MIMO schemes.
Simulation results also revealed large BER gains over the latter alternative, due to the higher coding gains afforded by the new method.
Since the proposed NGS requires shorter reference signals, it is expected to be suitable for low-overhead LPD communications in high-mobility scenarios.

\footnotesize{
\bibliographystyle{IEEEtranURLandMonthDiactivated}
\bibliography{main}
}

\end{document}